%% file: paper.tex
\documentstyle[12pt,epsfig]{article}

\newcommand{\ppp}[1]{%
        \setbox0=\hbox{#1}%
        \kern-.02em\copy0\kern-\wd0
        \kern+.04em\copy0\kern-\wd0
        \kern-.02em\raise.0217em\box0}
\newcommand{\vek}[1]{
         \mathchoice{\mbox{\boldmath$#1$}}%
        {\mbox{\boldmath$#1$}}%
        {\ppp{$\scriptstyle#1$}}%
        {\ppp{$\scriptscriptstyle#1$}}}

\newcommand{\lsim}{$\raisebox{-0.8ex} {$\stackrel{\textstyle <}{\sim}$}$}

\newcommand{\xpom}{x_{_{\rm I\!P}}}
\newcommand{\pom}{\rm I\!P}

\newcommand{\areg}{\alpha_{_{\rm I\!R}}}
\newcommand{\apom}{\alpha_{_{\rm I\!P}}}

\begin{document}  
\begin{titlepage}
\vspace*{-2cm}
\begin{flushright}
\bf 
TUM/T39-00--19
\end{flushright}

\bigskip 

\begin{center}
{\large\bf Nuclear Shadowing  
and the 
\\[0.3cm] 
Optics of Hadronic Fluctuations}\footnote{Work supported in
    part by BMBF}

\vspace{2.cm}

{\large G.~Piller$^{(a)}$, M.~V\"anttinen$^{(a)}$, 
L.~Mankiewicz$^{(a,b)}$ and W.~Weise$^{(a)}$ 
}
\date{\today{}}

\vspace{2.cm}

${(a)}$ Physik-Department, Technische Universit\"{a}t M\"{u}nchen,
D-85747 Garching, Germany 
\\
${(b)}$ N. Copernicus Astronomical Center, ul. Bartycka 18,
 PL--00-716 Warsaw, Poland

\vspace*{5cm}

{\bf Abstract}
\bigskip

\begin{minipage}{15cm}
A coordinate space description of shadowing in deep-inelastic lepton-nucleus
scattering is presented. The picture in the laboratory frame is that of
quark-gluon fluctuations of the high-energy virtual photon, propagating
coherently over large light-cone distances in the nuclear medium. We discuss
the detailed dependence of the coherence effects on the invariant mass of the
fluctuation. We comment on the issue of possible saturation in the shadowing
effects at very small Bjorken-$x$.

\end{minipage}

\end{center} 

\vspace*{1.cm}

%{\sl \centerline {To be published in Z. Phys. C.}}

\vspace*{2.cm}

%\noindent $^{*}$) Work supported in part by grants from BMFT  and ARC.

\end{titlepage}

%%%%%%%%%%%%%%%%%%

\section{Introduction: shadowing, diffraction and coherence length}

High precision data on shadowing effects  
in deep-inelastic lepton-nucleus  scattering have become available 
from measurements at CERN (NMC) and FNAL (E665) (for recent reviews see 
e.g. \cite{GP_rep,GST_rep,Arneodo_rep}).
Shadowing appears at small values of the Bjorken variable,
$x = Q^2/(2 M \nu) < 0.1$,
as a significant reduction of nuclear structure functions 
$F_2^{ A}$ compared to the free nucleon one  
($q^{\mu} = (\nu,\vek  q)$ is the four-momentum of the exchanged 
virtual photon with $Q^2 = - q^2$, and $M$ is the nucleon mass). 
%For illustration we show a 
%compilation of  data on the structure function ratio 
%$F_2^A/F_2^d$ for various nuclei in Fig. \ref{fig:RAd_shad1}. 
%
%%%%%%%%%%%%%%%%%%%%%%%%%%%%%%%%%%%%%%%%%%%%%%%%%%%%55
%\begin{figure}[h]
%\bigskip
%\begin{center} 
%\epsfig{file=nmc_logx.eps,height=80mm,width=70mm,angle=-90}
%\end{center}
%\caption[...]{
%NMC data \cite{Amaudruz:1995tq} for the  
%structure function ratio $F_2^{A} / F_2^{d}$ for  
%$^{4}$He, $^{12}$C, and $^{40}$Ca.  
%}
%\label{fig:RAd_shad1}
%\end{figure}
%%%%%%%%%%%%%%%%%%%%%%%%%%%%%%%%%%%%%%%%%%%%%%%%%%%%
%

A closely related process, 
diffractive photo-  and leptoproduction 
of hadrons from free nucleons, has been investigated in great 
detail lately at the HERA collider 
(for a review see e.g. \cite{Abramowicz:1998ii}). 
The connection between shadowing and diffraction 
has been observed already in Ref.\cite{Gribov:1970}: both processes  
are driven by the scattering of  hadronic components 
present in the interacting high-energy (virtual) photon. 
While diffraction reveals the mass spectrum of these hadronic 
fluctuations, 
shadowing is a direct measure of their coherent interaction 
with the nuclear medium. Nuclear shadowing in deep-inelastic lepton scattering 
can therefore be viewed as the  
{\em optics of hadronic (quark-antiquark plus gluon) components of the photon 
in the nuclear medium}. 
It provides a tool for investigating the hadronic structure and properties of high-energy 
(virtual) photons.

For the following discussion we choose the laboratory frame, 
where the target is at rest, and take  
the photon momentum $\vek q$ in the $\hat z$-direction. 
The nuclear structure function $F_2^{A}$, or equivalently, the photon-nucleus cross section 
$\sigma_{\gamma^* A}  = (4 \pi^2 \alpha/Q^2) A \,F_2^{A}$ 
can be separated into a piece which accounts for the incoherent 
scattering from individual nucleons, and a correction due to the 
coherent interaction with several nucleons, 
$\sigma_{\gamma^* A} = {A} 
\sigma_{\gamma^* N} + \delta \sigma_{\gamma^* A}$. 
The dominant contribution to coherent multiple scattering 
is described by the well known relation 
\cite{Gribov:1970,Kondratyuk:1973jept}: 
\begin{eqnarray} \label{eq:shad_diff}
&&\delta \sigma_{\gamma^*  A} = 
- 8 \pi \int_{4 m_{\pi}^2}^{W^2} dM_{X}^2
\int d^2 b \int_{-\infty}^{+\infty} dz_1 
\int_{z_1}^{+\infty} dz_2 \, 
\rho_{A}^{(2)}(\vek  b, z_1, z_2) 
\cos\left[(z_1 - z_2)/\lambda_X\right] 
\nonumber \\
&&\hspace*{3cm}
\times
\left.\frac{d^2\sigma_{\gamma^* N}^{diff}}{dM^2_{X}dt}
\right|_{t\approx 0} 
\,\exp\left[-\frac{\sigma_{{XN}}}{2}\int_{z_1}^{z_2} 
dz \, \rho_{A}(\vek  b, z)\right].
\end{eqnarray}
A state $X$ with invariant mass $M_{X}$ is produced 
diffractively in the process $\gamma^* N \rightarrow {X N}$, 
with the nucleon located at $(\vek  b,z_1)$. 
The upper limit of $M_{X}$ is determined by the available $\gamma^* N$ 
center-of-mass energy $W$. 
The hadronic excitation propagates at fixed impact parameter $\vek  b$ 
and scatters from  a second nucleon at $z_2$. 
The underlying basic mechanism, i.e. diffractive production from a single 
nucleon, is described by the cross section
${d^2\sigma_{\gamma^* N}^{diff}}/{dM_{X}dt}$ 
taken in the forward direction with $t =(p-p')^2\approx 0$, 
where $p$ and $p'$ are the four-momenta of the 
active nucleon before and after the diffractive  interaction.
The nuclear two-particle density, 
$\rho_{A}^{(2)} (\vek  b, z_1,z_2)$
accounts for the probability to find two nucleons in the target at 
the same impact parameter. 
The exponent in (\ref{eq:shad_diff}) approximates additional interactions of the diffractively produced hadronic states
while they propagate from $z_1$ to $z_2$. 
The strength of these higher order multiple scattering 
contributions is described by the effective cross section 
$\sigma_{{XN}}$. 
Finally, the $\cos\left[(z_1 - z_2)/\lambda\right]$ 
factor implies that only those diffractively produced hadrons 
which have a longitudinal propagation length (coherence length)
\begin{equation} \label{eq:coherence_length}
\lambda_{X} = \frac{2 \nu}{Q^2 + M_{X}^2} = 
\frac{1}{M x}\left(\frac{Q^2}{Q^2 + M_{X}^2}\right) 
\end{equation}
larger than the average nucleon-nucleon  separation 
($d \simeq 2 fm)$ in the target nucleus, 
can contribute significantly to coherent multiple scattering.

Shadowing should therefore start as soon as 
\begin{equation}
\lambda_X > d \simeq 2\;\rm{fm} \, .
\label{shadcond}
\end{equation}
From Eq.(\ref{eq:coherence_length}) one finds that 
at $Q^2 \gg 1$ GeV$^2$  the condition (\ref{shadcond}) is met 
at $x \, \lsim \, 0.1$ in accordance with the measured effect. Since $\lambda_{X}$
decreases with increasing $M_{X}$,  
low mass excitations with $M_{X} \,\lsim \,1$ GeV
are relevant for the onset of shadowing. When $x$ decreases far below $0.1$, shadowing increases strongly.   
This behavior is mainly caused by  the diffractive production and coherent 
multiple scattering of hadronic states with masses $M_{   X} >  1$ GeV. At $x
\ll 0.1$ it is also the energy dependence of   
the diffractive production cross section and of the hadron-nucleon cross 
section $\sigma_{{X N}}$ which influences 
the $x$-dependence of shadowing.  

To illustrate this behavior, consider the shadowing ratio 
$R_{A} = {\sigma_{\gamma^* {A}}}/{A \sigma_{\gamma^* {N}}} 
= 1 - \delta\sigma_{\gamma^* A}/{A \sigma_{\gamma^* {N}}}$,  
parametrized as:
\begin{equation}  \label{eq:RA-1}
R_{A} - 1 = 
- c \, \left(\frac{1}{x}\right)^{\varepsilon},
\end{equation}
with a constant $c$ at small $Q^2$ where data are actually taken, 
and a characteristic exponent $\epsilon$. 
At asymptotically large energies Regge phenomenology 
suggests $\varepsilon \simeq 0.1$ \cite{Donnachie:1992ny,Collins:1977jy}. 
\footnote{
At the typical average center of mass energies 
$\overline  W < 25$ GeV 
used in experiments at CERN and FNAL, 
a somewhat stronger energy dependence is expected 
through the kinematic restriction to diffractively produced 
hadronic states with masses $M_{ X} < W$.}
%%%%%%%%%%%%%%%%%%%%%%%%%%%%%%%%%%%%%%%%%%%%%%%%%%%%55
\begin{figure}[h]
\bigskip
\begin{center} 
\epsfig{file=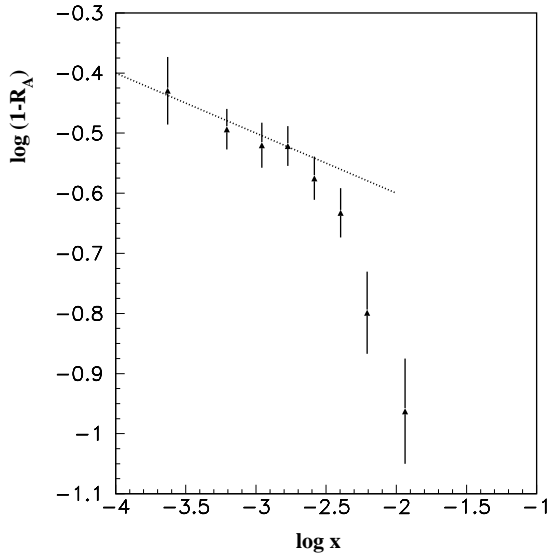,height=80mm}
\end{center}
\caption[...]{
The quantity $\log \left( 1 - R_{A} \right)$  
as a function of $\log x$ for  
data taken on Pb \cite{Adams:1995is}. 
The dashed line corresponds to the asymptotic energy dependence 
(\ref{eq:log_est}) with $\varepsilon = 0.1$. 
}
\label{fig:logR}
\end{figure}
%%%%%%%%%%%%%%%%%%%%%%%%%%%%%%%%%%%%%%%%%%%%%%%%%%%%
%
In Fig.\ref{fig:logR} we show the quantity
\begin{equation}  \label{eq:log_est}
\log \left( 1 - R_{ A} \right) = \log c  - \varepsilon \, \log x, 
\end{equation}
plotted versus $\log x$ in comparison with data taken on Pb at 
small $Q^2$. 
This plot confirms that, for $x < 3 \cdot 10^{-3}$, 
the shadowing effect indeed 
approaches the high-energy behavior expected from the 
Regge limit of diffractive production.  
Deviations from this asymptotic behavior at larger values of $x$ indicate 
how shadowing gradually builds up as the coherence length 
$\lambda_X \propto x^{-1}$ starts to exceed nuclear length scales for 
diffractively produced states of low mass. At sufficiently high energy or small
$x$, the coherence length becomes comparable to nuclear dimensions even for 
heavy hadronic intermediate states. 
%Once a major fraction of diffractively produced states 
%contribute to shadowing it starts to approach its asymptotic 
%high-energy behavior. 
Then shadowing starts to approach its asymptotic 
high-energy behavior. 

Note that this  asymptotic behavior of shadowing sets in when the 
coherence lengths $\lambda_{X}$ of low mass 
hadronic fluctuations exceed by far the diameter of the nucleus.
For example, at $x = 3\cdot 10^{-3}$ and $Q^2 \simeq 0.7$ GeV$^2$   
which corresponds to the onset of the asymptotic behavior in 
Fig.\ref{fig:logR},  the $\rho$ meson coherence length becomes 
$\lambda_{\rho} \simeq 36$ fm. 
The influence of the nuclear size on shadowing is not apparent, and we wish to
understand why this is so.

In the following we present a detailed analysis of the role 
played by characteristic length scales of nuclei in 
high energy deep-inelastic scattering.
In Section 2 we recall the basic features of nucleon and nuclear 
structure functions in coordinate space.
The influence of nuclear scales on the coherent interaction of 
hadronic states present in the diffractive mass spectrum of the 
photon is discussed in Section 3. 

\section{Structure functions in coordinate space}

The space-time evolution of   
nuclear deep-inelastic scattering is viewed most instructively 
in coordinate space \cite{Hoyer:1996nr,Vanttinen:1998iz}
where one can directly compare the characteristic distances involved 
in deep-inelastic scattering with scales provided by the 
nuclear target. 

The coordinate-space structure function ${\cal F}_2$ is related 
to the momentum-space structure function $F_2$ 
by:
\begin{equation} \label{eq:F2_Ioffe} 
{\cal F}_2 (l,Q^2) = \int_0^1 \frac{dx}{x}\, 
F_2(x,Q^2) \,\sin\left(l M x \right). 
\end{equation}
In the laboratory frame with the $\hat z$-axis along the 
direction of the incident photon momentum, 
the Fourier transform (\ref{eq:F2_Ioffe}) 
projects out contributions from the light-cone, i.e. 
from time and space intervals along the direction of the photon. We have $t+z =
2 l$, and the longitudinal distances 
involved in the process are $z \simeq t \simeq l$.

%%%%%%%%%%%%%%%%%%%%%%%%%%%%%%%%%%%%%%%%%%%%%%%%%%%%55
\begin{figure}[h]
\bigskip
\begin{center} 
\epsfig{file=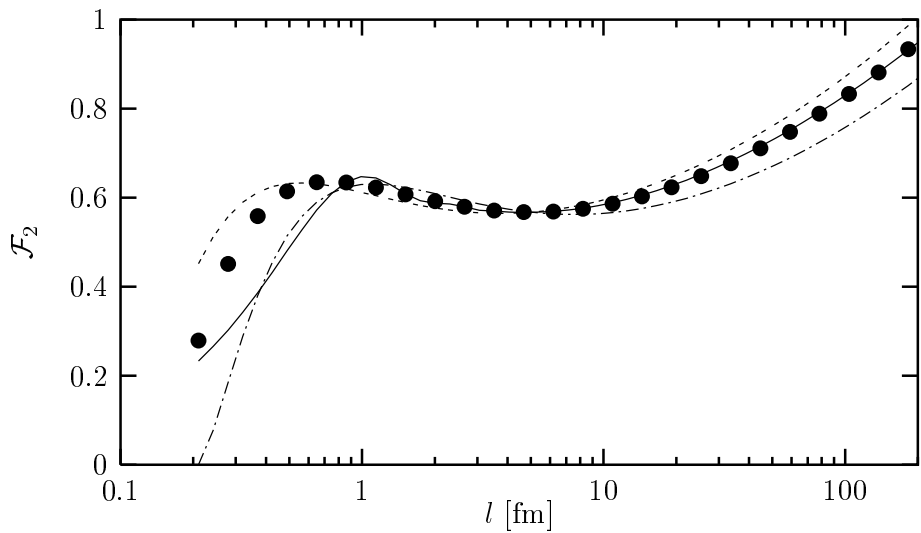,height=70mm}
\end{center}
\caption[...]
{Full line: empirical proton structure function ${\cal F}_2$ in coordinate
space for 
$Q^2 = 2.25$ GeV$^2$ as obtained from the parametrization of 
Ref.\cite{Abramowicz:1997ms}. The full dots correspond to the approximate
relation  ${\cal F}_2(l) \sim F_2(x = 1/(1.5\, M l))$. 
The dashed and dash-dotted  curves correspond  to $F_2(x= 1/(2 M l))$ 
and $F_2(x=(1/M l))$, respectively. The approximate coordinate-space 
distributions are normalized to the exact result at $l=5$ fm.
}
\label{fig:f2Ncoord}
\end{figure}
%%%%%%%%%%%%%%%%%%%%%%%%%%%%%%%%%%%%%%%%%%%%%%%%%%%%
%
We show in Fig.\ref{fig:f2Ncoord} the proton structure function 
${\cal F}_2$ at  $Q^2 = 2.25$ GeV$^2$ as obtained from the parametrization 
of Ref.\cite{Abramowicz:1997ms}.
One observes that ${\cal F}_2$ rises at small $l$, develops 
a plateau at $l\simeq 1$ fm, and increases further at large $l$. 
At $l < 1$ fm ${\cal F}_2$  is determined 
by average properties of the corresponding momentum-space structure  
function $F_2$, as expressed by its first few moments 
\cite{Mankiewicz:1996ep,Weigl:1996ii}. 
For example, the derivative of ${\cal F}_2$ at $l=0$ 
reflects the fraction of the nucleon light-cone momentum carried by quarks. 
At large distances, $l > 2$ fm,  ${\cal F}_2$ is 
governed by the small-$x$ part of $F_2$ and $l \sim 1/Mx$. 
As an example, we find approximately
${\cal F}_2(l) \sim F_2((x = 1/(1.5\, M l))$ 
for $Q^2 = 0.5$ -- $6$ GeV$^2$ as illustrated in  
Fig.\ref{fig:f2Ncoord}. 

Evidently, the interacting photon-nucleon system at high energy stretches over
distances very large compared to what one usually associates with the ''size''
of the nucleon. 
The fact that  ${\cal F}_2$ extends over such large distances 
has a natural interpretation in the laboratory frame:  
at correlation lengths $l$ much larger than the nucleon 
size, ${\cal F}_2$ reflects primarily the partonic structure of 
the interacting photon.

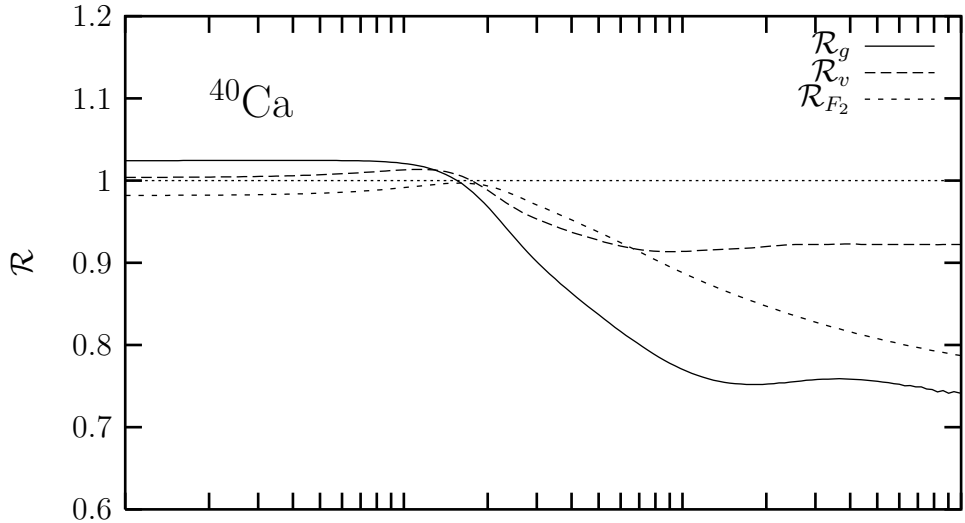
\begin{figure}
\input{cSAEF.Ca.tex}
\caption{Coordinate-space ratio ${\cal R}_{F_2} = {\cal F}_2^{Ca}/{\cal
F}_2^{N}$ at $Q^2 = 4$ GeV$^2$ for  
the $F_2$ structure function of $^{40}$Ca as taken from 
Ref.\cite{Vanttinen:1998iz}. 
Also shown are the ratios for the gluon distributions, ${\cal R}_g$ and the 
valence-quark distributions, ${\cal R}_v$, as discussed in \cite{Vanttinen:1998iz}.
Note that the nucleon structure functions are normalized per nucleon.}
\label{fig:coordspace-Ca-Pb}
\end{figure}

Consider now deep-inelastic scattering from nuclei. 
The gross features of the observed nuclear effects 
can be interpreted in a surprisingly simple way when 
considered  in coordinate-space.  
Two different kinematic regions can be distinguished 
immediately:
\begin{itemize}

\item
At distances $l<d$, smaller than the average spacing between two nucleons, $d
\sim 2$ fm, the virtual photon interacts 
incoherently 
with the constituents bound in the target nucleus.

\item
At larger distances, $l>d$, several nucleons can participate 
in the interaction. Deviations of nuclear structure functions 
from the one of the free nucleon are then expected to come from coherent
scattering on at least  
two nucleons. 

\end{itemize}
Fig.\ref{fig:coordspace-Ca-Pb} shows the ratio of the calcium and free nucleon 
coordinate-space structure functions, 
${\cal R}_{F_2} = {\cal F}_2^{Ca}/{\cal F}_2^{N}$, as obtained  
from a fit to empirical nuclear momentum-space structure functions 
\cite{Vanttinen:1998iz}. 
One observes  a clear separation of nuclear effects at small 
and large distances $l$. 
At $l > 2$ fm a strong depletion of the nuclear 
structure function is found: the virtual photon behaves 
like a beam of quarks and gluons which scatters coherently 
from several nucleons in the target causing nuclear shadowing. 
On the other hand, at $l < 2$ fm nuclear effects are small, 
indicating that the intrinsic structure of bound nucleons 
is not much changed in the nuclear environment.

\section{Shadowing and nuclear scales}

While the average separation of nucleons in nuclei evidently plays a 
prominent  role in the nuclear structure functions, 
the size of the 
target nucleus itself seems not to be important: shadowing 
continues to increase for distances much larger than the 
nuclear diameter.

The reason for this behavior is found in the fact that 
the interacting virtual photon couples to a spectrum of 
hadronic fluctuations which come  with different propagation lengths.
To investigate this issue further we consider in the following 
the coordinate-space representation of the shadowing correction 
$\delta F_2^{A} = F_2^{A} - F_2^{N}$: 
\begin{equation} \label{eq:delta_F2_Ioffe} 
\delta {\cal F}_2^{   A} (l,Q^2) = \int_0^1 \frac{dx}{x}\, 
\delta F_2^{  A}(x,Q^2) \,\sin\left(l M x \right). 
\end{equation}
We are interested in the influence of the 
invariant mass $M_{ X}$ of the diffractively 
produced hadronic states, as it appears in the coherence length
(\ref{eq:coherence_length}). 
For this purpose, consider first a simple ansatz  
for the diffractive production cross section which 
determines the shadowing correction in Eq.(\ref{eq:shad_diff}). 
%We describe the diffractive production of a hadronic state 
%$V$ with invariant mass $m_{ V}$ in the spirit of 
Suppose that a single, ''vector meson'' - like state $X = V$ with invariant
mass $m_V$ contributes, and that its diffractive production cross-section has
the form suggested by
vector meson dominance 
\cite{Bauer:1978iq}:
\begin{equation} \label{eq:vmd_shad}
\left.\frac{d^2\sigma_{\gamma^*   N}^{diff}}{dM^2_{  X}dt}
\right|_{t\approx 0} 
= \frac{\alpha}{4} 
\frac{m_{ V}^2\,\sigma_{{VN}}^2}
{(m_{ V}^2 + Q^2)^2} 
\left(\frac{m_{ V}}{g_{ V}} \right)^2 
\delta\left(m_{ V}^2 - M_{ X}^2\right). 
\end{equation}
In this schematic discussion we use a constant hadron-nucleon cross section,
say $\sigma_{{VN}} \approx 20$ mb, and assume that the dimensionless coupling
$g_V$ scales roughly with $m_V$.  
%\footnote{This choice mimiks the aligned-jet-model  
%ansatz for deep-inelastic scattering at small $x$ \cite{...}. 
%Here the phase space of quark-antiquark fluctuations in the photon 
%with a hadronic interaction cross section decreases with increasing 
%mass $m_{ V}$.}
\footnote{For the following discussion the detailed
numerical values of $\sigma_{VN}$ and 
$g_V$ are actually not relevant.}
The two-body nuclear density in (\ref{eq:shad_diff}) is expressed as  
a product of one-body densities multiplied with 
a nucleon-nucleon correlation function, 
$\rho_{ A}^{(2)} (\vek b,z_1,z_2) 
= \rho_{ A}(\vek b,z_1) \rho_{ A} (\vek b,z_2) 
C(z_1 - z_2)$. 
The one-body densities $\rho_{ A}(\vek b,z_1)$ are 
parametrized by a square well, normalized to the nuclear 
mass number $ A$. 
For the correlation function we take a simple step function
$C(z) = K \Theta(|z| - z_0)$ with $z_0 = 0.8$ fm \cite{corr} and 
$K$ fixed by the normalization $\int d^3 r_1 d^3 \,r_2
\rho(\vek r_1,\vek r_2) = A^2$.

%%%%%%%%%%%%%%%%%%%%%%%%%%%%%%%%%%%%%%%%%%%%%%%%%%%%55
\begin{figure}[h]
\bigskip
\begin{center} 
\epsfig{file=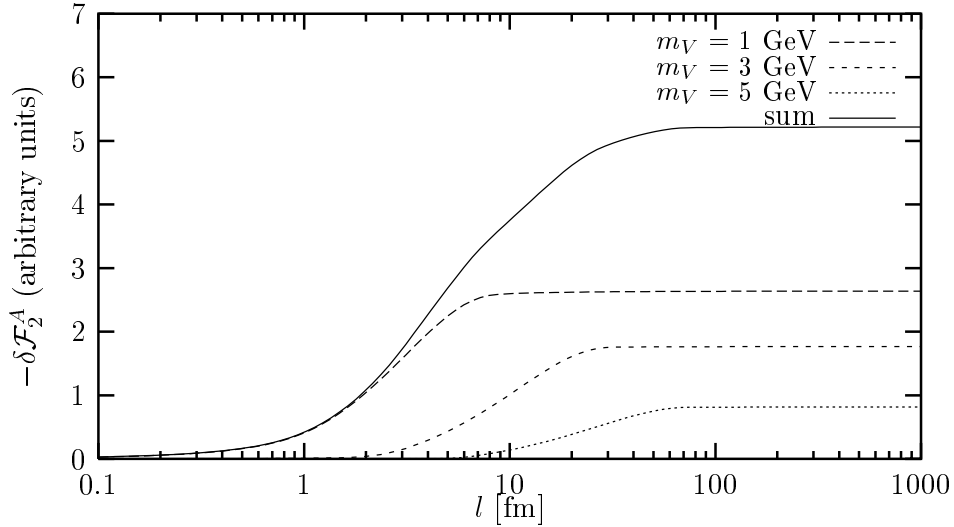,height=70mm}
\end{center}
\caption[...]{The shadowing correction 
$\delta {\cal F}_2^{A}$  
of Eq.(\ref{eq:delta_F2_Ioffe}) derived from the
schematic 
model (\ref{eq:vmd_shad}) for $^{12}$C,  
calculated at  $Q^2 = 4$ GeV$^2$ using three different values for the 
invariant mass $m_V$ of the interacting hadronic state. The solid curve is the
sum of all three contributions.
}
\label{fig:R_toy}
\end{figure}
%%%%%%%%%%%%%%%%%%%%%%%%%%%%%%%%%%%%%%%%%%%%%%%%%%%%
%
Fig.\ref{fig:R_toy} shows the shadowing correction 
$\delta {\cal F}_2^{   A}$ in coordinate space for $^{12}$C 
at $Q^2 = 4$ GeV$^2$ using different values 
for the invariant mass $m_V$ of the interacting hadronic state.  
The onset of shadowing, as well as its saturation at large distances $l$,  
shows a significant dependence on $m_V$.
This can easily be explained by recalling Eq.(\ref{eq:coherence_length}). 
At small values of Bjorken-$x$ the longitudinal 
interaction length involved in deep-inelastic scattering 
is given approximately 
by $l\simeq  2/(3 Mx)$ as illustrated in Fig.\ref{fig:f2Ncoord}. 
In combination with  Eq.(\ref{eq:coherence_length}) 
one finds that the coherence length 
$\lambda$ of a hadronic fluctuation with mass $m_{ V}$ 
exceeds a given distance $\Delta$ when: 
\begin{equation} \label{eq:cond_sat}
l > \frac{2 \Delta}{3} \left(1 + \frac{m_V^2}{Q^2}\right).
\end{equation}
The onset of shadowing occurs for distances close to the 
average nucleon-nucleon separation in nuclei, $\Delta \simeq d \simeq  2$ fm. 
For $m_V \simeq 1$ GeV and $Q^2 \simeq 2$ GeV$^2$ the condition
(\ref{eq:cond_sat}) is   
satisfied at $l \simeq 2$ fm. 
For heavier states Eq.(\ref{eq:cond_sat}) implies larger values of $l$, in agreement with Fig.\ref{fig:R_toy}.

At large distances shadowing saturates for each given $m_V$ and approaches a constant value. 
This behavior is expected once the coherence length gets 
larger than the nuclear diameter,
i.e. $\lambda > 2 R_{ A}$.
As $m_V$ increases,
larger values of $l$ 
are needed for both the onset and the saturation of shadowing. 
With $\Delta = 2 R_{^{12} C} \approx 5$ fm, the
condition (\ref{eq:cond_sat}) reproduces the 
values of $l$ where saturation is observed in Fig.\ref{fig:R_toy}. 

In summary, the shadowing correction caused 
by a hadronic state with fixed invariant mass $m_{ V}$ 
reveals two characteristic nuclear scales, 
namely the average distance between two nucleons and 
the nuclear diameter. 
However, summing over contributions 
from hadronic fluctuations with different 
masses masks the role played by the nuclear diameter.

We have just demonstrated these features of coherent nuclear shadowing using a
simple schematic model. This can of course also be 
verified within a more realistic description of 
diffractive photon-nucleon  scattering. 
Consider a parametrization \cite{Melnitchouk:1993vc} 
for the diffractive cross section,
based on Regge phenomenology. 
The differential production cross section 
for hadronic states with invariant mass 
$M_X > 1.2$ GeV is written in factorized form: 
\begin{equation}
\left.\frac{d\sigma^{diff}}{d \xpom dt}\right|_{t =0} = 
W^2 \frac{d\sigma^{diff}}{d M_X^2 dt} = 
\frac{Q^2}{4 \pi^2 \alpha} \, f_{\pom/N}(\xpom) F_{2\pom}(x/\xpom,Q^2),  
\end{equation}
where $F_{2\pom}$ is commonly interpreted as the ``structure function'' 
of the pomeron, and 
\begin{equation}
f_{\pom/N}(\xpom) = \frac{\sigma_{pp}}{16 \pi} 
\frac{1}{\xpom^{2 \apom -1}}
\end{equation}
describes its distribution in the nucleon, 
with $\xpom = x\left(1 + \frac{M_{X}}{Q^2}\right)$ 
the fraction of the nucleon momentum carried by 
the pomeron.
The intercept of the pomeron is taken to be $\apom = 1.08$ 
\cite{Donnachie:1992ny}.
The proton-proton cross section 
(due to pomeron exchange)  
is fixed at $\sigma_{pp}=40$ mb \cite{Melnitchouk:1993vc}.

The pomeron structure function is split into a quark-antiquark 
component and a contribution due to triple pomeron exchange, 
$F_{2\pom} = F_{2\pom}^{(q\bar q)}+F_{2\pom}^{(3 \pom)}$, 
where 
\begin{eqnarray}
F_{2\pom}^{(q\bar q)} &=& \frac{8(5 + \lambda_s) \beta_0^2}{3 \sigma_{pp}}
\frac{N_{sea} Q^2}{Q^2 + Q_0^2} \,\beta (1-\beta),  
\\
F_{2\pom}^{(3 \pom)} &=& 
\frac{g_{3\pom}}{\sqrt \sigma_{pp}} \,
F_{2}^{N\,sea}(\beta,Q^2). 
\end{eqnarray}
Here $\beta = x/\xpom$ is the fraction of the pomeron's momentum 
carried by the struck quark. 
Following \cite{Melnitchouk:1993vc} we use  
$\beta_0^2 = 3.4$ GeV$^{-2}$, $\lambda_s = 0.5$, 
$N_{sea} = 0.17$, $Q_0^2 = 0.485$ GeV$^2$, 
and $g_{3\pom} = 0.364$ mb$^{1/2}$. 
These parameters give a good description of diffractive photoproduction on
nuclei, when combined with a cross section $\sigma_{XN} \simeq 20$ mb for the
re-scattering of quark-gluon configuration in a multiple scattering expansion.
The sea quark contribution $F_{2}^{N, sea}$ to the nucleon structure 
function $F_2$ has been extracted from the parametrizations 
of Refs.\cite{Abramowicz:1997ms}. 

The contributions of the light vector mesons $\rho$, $\omega$, 
and $\phi$ are described by the vector meson 
dominance expression (\ref{eq:vmd_shad}). 
The vector meson coupling constants are fixed at $g_{\rho} = 5.0$, 
$g_{\phi} = 16.3$, and $g_{\omega} = 13.2$. 
The vector meson-nucleon cross sections 
are parametrized as: 
$\sigma_{\rho N} = 
\sigma_{\omega N} = 13.6 s^{\apom -1} + 31.8 s^{\areg-1}$ 
and  
$\sigma_{\phi N} = 10.0 s^{\apom -1} + 1.5  s^{\areg-1}$, 
with $s = W^2$ and $\areg = 0.55$.

%In Refs.\cite{Melnitchouk:1993vc} the above description of diffractive 
%photoproduction in
%photon-nucleon scattering has been used to calculate the 
%multiple scattering correction according from Eq.(\ref{eq:shad_diff}). 
%For the re-scattering cross section for quark-gluon configurations 
%with large mass  a typical hadronic  value, $\sigma_{XN} = 20$ mb,  has been taken.

%%%%%%%%%%%%%%%%%%%%%%%%%%%%%%%%%%%%%%%%%%%%%%%%%%%%55
\begin{figure}[h]
\bigskip
\begin{center} 
\epsfig{file=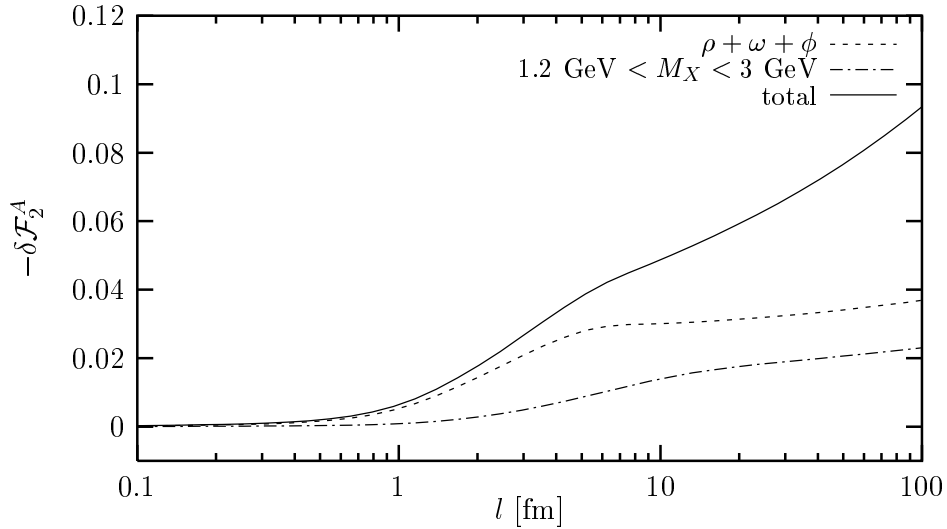,height=70mm}
\end{center}
\caption[...]{The shadowing correction 
$\delta F_{2}^A$  in coordinate space (\ref{eq:delta_F2_Ioffe}) as obtained 
from the parametrization of the diffractive production cross 
of Ref.\cite{Melnitchouk:1993vc} taken at $Q^2 = 4$ GeV$^2$. Contributions 
from vector mesons and hadronic states  with invariant mass 
$1.2$ GeV $< M_X < 3$ GeV are shown separately.
}
\label{fig:shad_real}
\end{figure}
%%%%%%%%%%%%%%%%%%%%%%%%%%%%%%%%%%%%%%%%%%%%%%%%%%%%
%
With these ingredients data on nuclear 
shadowing from CERN \cite{Amaudruz:1995tq} and FNAL \cite{Adams:1992nf} can be
reproduced satisfactorily. 
In Fig.\ref{fig:shad_real} we show the resulting shadowing correction 
$\delta F_{2}^A$  in coordinate space, Eq.(\ref{eq:delta_F2_Ioffe}),
taken at $Q^2 = 4$ GeV$^2$. 
Contributions from light vector mesons and  from hadronic states with
intermediate invariant 
mass,
$1.2$ GeV $< M_X < 3$ GeV, are displayed separately. 
We observe the same features as in our previous toy-model results of 
Fig.\ref{fig:R_toy}:
a specific hadronic state in the diffractive mass spectrum of the photon 
contributes significantly to shadowing as soon as its 
coherence length $\lambda_X$ exceeds the average nucleon-nucleon distance.  
Saturation occurs when $\lambda_X$ becomes larger than the nuclear diameter, or
equivalently, for
\begin{equation} \label{eq:sat_cond}
l \stackrel{>}{\sim} \frac{4 R_A}{3} \left( 1 + \frac{M_X^2}{Q^2} \right)
\end{equation}
Then the weak dependence of shadowing on $l$ reflects the weak  
energy dependence of the diffractive photon-nucleon cross section. 
However, after summing over the shadowing corrections from all hadronic 
states with masses in the entire range $M_X < W$ the saturation
condition (\ref{eq:sat_cond}) is dispersed over a broad $l$-interval and the
dependence on the nuclear size gets hidden, leaving the average nucleon-nucleon
distance as the only remaining ''visible'' scale.

\section{Summary}

Shadowing in deep-inelastic lepton-nucleus scattering at small Bjorken-$x$ can
be viewed as an effect related to the coherent propagation of quark-gluon
fluctuations of the virtual photon in the nuclear medium, hence the close
analogy to notions familiar from wave optics.

This phenomenon is best illustrated in coordinate (rather then momentum) space.
At $x << 0.1$, the quark-gluon fluctuations extend along the light-cone
over distances large compared to the size of the nucleus. The picture in the
laboratory frame is that of a ''beam'' of quark-antiquark pairs and gluons
passing through the target nucleus.

The coherence (or propagation) length of each spectral component of this
''beam'' is governed by its invariant mass. Low mass components propagate over
the largest distances and behave as vector mesons. Higher mass components require
very high photon energies (at fixed $Q^2$) or, equivalently, very small
Bjorken-$x$ in order to travel over large distances. These components are
properly treated within Regge phenomenology, as in high-energy diffractive
production. 

Each individual spectral component has a distinct shadowing behaviour as a
function of the longitudinal distance scale, with the shadowing effect
saturating at large distance when propagation length (at given invariant mass)
exceeds the nucleon diameter. However, when summing over all (low mass and high
mass) fluctuations, we expect that the nuclear radius disappears as a visible
parameter and saturation is not reached. This interesting phenomenon should
have observable consequences in nuclear structure functions at the smallest
possible values of Bjorken-$x$.

{\bf Acknowledgments}: 

This work was supported in part by
KBN grant 2~P03B~011~19.
 
\newpage
%%%%%%%%%%%%%%%%%%%%%%%%%

%%%%%%%%%%%%%%%%%%%%%%%%%%%%%%%%%%%%%%%%%%%%%%%%%%%%%%%%%%%%%%%%%%

\end{document}

%% file: cSAEF.Ca.tex
% GNUPLOT: LaTeX picture with Postscript
\setlength{\unitlength}{0.1bp}
\special{!
%!PS-Adobe-2.0 EPSF-2.0
%%Title: cSAEF.Ca.tex
%%Creator: gnuplot 3.5 (pre 3.6) patchlevel beta 347
%%CreationDate: Wed Aug 12 14:49:00 1998
%%DocumentFonts: 
%%BoundingBox: 0 0 360 216
%%Orientation: Landscape
%%EndComments
/gnudict 120 dict def
gnudict begin
/Color false def
/Solid false def
/gnulinewidth 5.000 def
/userlinewidth gnulinewidth def
/vshift -33 def
/dl {10 mul} def
/hpt_ 31.5 def
/vpt_ 31.5 def
/hpt hpt_ def
/vpt vpt_ def
/M {moveto} bind def
/L {lineto} bind def
/R {rmoveto} bind def
/V {rlineto} bind def
/vpt2 vpt 2 mul def
/hpt2 hpt 2 mul def
/Lshow { currentpoint stroke M
  0 vshift R show } def
/Rshow { currentpoint stroke M
  dup stringwidth pop neg vshift R show } def
/Cshow { currentpoint stroke M
  dup stringwidth pop -2 div vshift R show } def
/UP { dup vpt_ mul /vpt exch def hpt_ mul /hpt exch def
  /hpt2 hpt 2 mul def /vpt2 vpt 2 mul def } def
/DL { Color {setrgbcolor Solid {pop []} if 0 setdash }
 {pop pop pop Solid {pop []} if 0 setdash} ifelse } def
/BL { stroke gnulinewidth 2 mul setlinewidth } def
/AL { stroke gnulinewidth 2 div setlinewidth } def
/UL { gnulinewidth mul /userlinewidth exch def } def
/PL { stroke userlinewidth setlinewidth } def
/LTb { BL [] 0 0 0 DL } def
/LTa { AL [1 dl 2 dl] 0 setdash 0 0 0 setrgbcolor } def
/LT0 { PL [] 1 0 0 DL } def
/LT1 { PL [4 dl 2 dl] 0 1 0 DL } def
/LT2 { PL [2 dl 3 dl] 0 0 1 DL } def
/LT3 { PL [1 dl 1.5 dl] 1 0 1 DL } def
/LT4 { PL [5 dl 2 dl 1 dl 2 dl] 0 1 1 DL } def
/LT5 { PL [4 dl 3 dl 1 dl 3 dl] 1 1 0 DL } def
/LT6 { PL [2 dl 2 dl 2 dl 4 dl] 0 0 0 DL } def
/LT7 { PL [2 dl 2 dl 2 dl 2 dl 2 dl 4 dl] 1 0.3 0 DL } def
/LT8 { PL [2 dl 2 dl 2 dl 2 dl 2 dl 2 dl 2 dl 4 dl] 0.5 0.5 0.5 DL } def
/Pnt { stroke [] 0 setdash
   gsave 1 setlinecap M 0 0 V stroke grestore } def
/Dia { stroke [] 0 setdash 2 copy vpt add M
  hpt neg vpt neg V hpt vpt neg V
  hpt vpt V hpt neg vpt V closepath stroke
  Pnt } def
/Pls { stroke [] 0 setdash vpt sub M 0 vpt2 V
  currentpoint stroke M
  hpt neg vpt neg R hpt2 0 V stroke
  } def
/Box { stroke [] 0 setdash 2 copy exch hpt sub exch vpt add M
  0 vpt2 neg V hpt2 0 V 0 vpt2 V
  hpt2 neg 0 V closepath stroke
  Pnt } def
/Crs { stroke [] 0 setdash exch hpt sub exch vpt add M
  hpt2 vpt2 neg V currentpoint stroke M
  hpt2 neg 0 R hpt2 vpt2 V stroke } def
/TriU { stroke [] 0 setdash 2 copy vpt 1.12 mul add M
  hpt neg vpt -1.62 mul V
  hpt 2 mul 0 V
  hpt neg vpt 1.62 mul V closepath stroke
  Pnt  } def
/Star { 2 copy Pls Crs } def
/BoxF { stroke [] 0 setdash exch hpt sub exch vpt add M
  0 vpt2 neg V  hpt2 0 V  0 vpt2 V
  hpt2 neg 0 V  closepath fill } def
/TriUF { stroke [] 0 setdash vpt 1.12 mul add M
  hpt neg vpt -1.62 mul V
  hpt 2 mul 0 V
  hpt neg vpt 1.62 mul V closepath fill } def
/TriD { stroke [] 0 setdash 2 copy vpt 1.12 mul sub M
  hpt neg vpt 1.62 mul V
  hpt 2 mul 0 V
  hpt neg vpt -1.62 mul V closepath stroke
  Pnt  } def
/TriDF { stroke [] 0 setdash vpt 1.12 mul sub M
  hpt neg vpt 1.62 mul V
  hpt 2 mul 0 V
  hpt neg vpt -1.62 mul V closepath fill} def
/DiaF { stroke [] 0 setdash vpt add M
  hpt neg vpt neg V hpt vpt neg V
  hpt vpt V hpt neg vpt V closepath fill } def
/Pent { stroke [] 0 setdash 2 copy gsave
  translate 0 hpt M 4 {72 rotate 0 hpt L} repeat
  closepath stroke grestore Pnt } def
/PentF { stroke [] 0 setdash gsave
  translate 0 hpt M 4 {72 rotate 0 hpt L} repeat
  closepath fill grestore } def
/Circle { stroke [] 0 setdash 2 copy
  hpt 0 360 arc stroke Pnt } def
/CircleF { stroke [] 0 setdash hpt 0 360 arc fill } def
/C0 { BL [] 0 setdash 2 copy moveto vpt 90 450  arc } bind def
/C1 { BL [] 0 setdash 2 copy        moveto
       2 copy  vpt 0 90 arc closepath fill
               vpt 0 360 arc closepath } bind def
/C2 { BL [] 0 setdash 2 copy moveto
       2 copy  vpt 90 180 arc closepath fill
               vpt 0 360 arc closepath } bind def
/C3 { BL [] 0 setdash 2 copy moveto
       2 copy  vpt 0 180 arc closepath fill
               vpt 0 360 arc closepath } bind def
/C4 { BL [] 0 setdash 2 copy moveto
       2 copy  vpt 180 270 arc closepath fill
               vpt 0 360 arc closepath } bind def
/C5 { BL [] 0 setdash 2 copy moveto
       2 copy  vpt 0 90 arc
       2 copy moveto
       2 copy  vpt 180 270 arc closepath fill
               vpt 0 360 arc } bind def
/C6 { BL [] 0 setdash 2 copy moveto
      2 copy  vpt 90 270 arc closepath fill
              vpt 0 360 arc closepath } bind def
/C7 { BL [] 0 setdash 2 copy moveto
      2 copy  vpt 0 270 arc closepath fill
              vpt 0 360 arc closepath } bind def
/C8 { BL [] 0 setdash 2 copy moveto
      2 copy vpt 270 360 arc closepath fill
              vpt 0 360 arc closepath } bind def
/C9 { BL [] 0 setdash 2 copy moveto
      2 copy  vpt 270 450 arc closepath fill
              vpt 0 360 arc closepath } bind def
/C10 { BL [] 0 setdash 2 copy 2 copy moveto vpt 270 360 arc closepath fill
       2 copy moveto
       2 copy vpt 90 180 arc closepath fill
               vpt 0 360 arc closepath } bind def
/C11 { BL [] 0 setdash 2 copy moveto
       2 copy  vpt 0 180 arc closepath fill
       2 copy moveto
       2 copy  vpt 270 360 arc closepath fill
               vpt 0 360 arc closepath } bind def
/C12 { BL [] 0 setdash 2 copy moveto
       2 copy  vpt 180 360 arc closepath fill
               vpt 0 360 arc closepath } bind def
/C13 { BL [] 0 setdash  2 copy moveto
       2 copy  vpt 0 90 arc closepath fill
       2 copy moveto
       2 copy  vpt 180 360 arc closepath fill
               vpt 0 360 arc closepath } bind def
/C14 { BL [] 0 setdash 2 copy moveto
       2 copy  vpt 90 360 arc closepath fill
               vpt 0 360 arc } bind def
/C15 { BL [] 0 setdash 2 copy vpt 0 360 arc closepath fill
               vpt 0 360 arc closepath } bind def
/Rec   { newpath 4 2 roll moveto 1 index 0 rlineto 0 exch rlineto
       neg 0 rlineto closepath } bind def
/Square { dup Rec } bind def
/Bsquare { vpt sub exch vpt sub exch vpt2 Square } bind def
/S0 { BL [] 0 setdash 2 copy moveto 0 vpt rlineto BL Bsquare } bind def
/S1 { BL [] 0 setdash 2 copy vpt Square fill Bsquare } bind def
/S2 { BL [] 0 setdash 2 copy exch vpt sub exch vpt Square fill Bsquare } bind def
/S3 { BL [] 0 setdash 2 copy exch vpt sub exch vpt2 vpt Rec fill Bsquare } bind def
/S4 { BL [] 0 setdash 2 copy exch vpt sub exch vpt sub vpt Square fill Bsquare } bind def
/S5 { BL [] 0 setdash 2 copy 2 copy vpt Square fill
       exch vpt sub exch vpt sub vpt Square fill Bsquare } bind def
/S6 { BL [] 0 setdash 2 copy exch vpt sub exch vpt sub vpt vpt2 Rec fill Bsquare } bind def
/S7 { BL [] 0 setdash 2 copy exch vpt sub exch vpt sub vpt vpt2 Rec fill
       2 copy vpt Square fill
       Bsquare } bind def
/S8 { BL [] 0 setdash 2 copy vpt sub vpt Square fill Bsquare } bind def
/S9 { BL [] 0 setdash 2 copy vpt sub vpt vpt2 Rec fill Bsquare } bind def
/S10 { BL [] 0 setdash 2 copy vpt sub vpt Square fill 2 copy exch vpt sub exch vpt Square fill
       Bsquare } bind def
/S11 { BL [] 0 setdash 2 copy vpt sub vpt Square fill 2 copy exch vpt sub exch vpt2 vpt Rec fill
       Bsquare } bind def
/S12 { BL [] 0 setdash 2 copy exch vpt sub exch vpt sub vpt2 vpt Rec fill Bsquare } bind def
/S13 { BL [] 0 setdash 2 copy exch vpt sub exch vpt sub vpt2 vpt Rec fill
       2 copy vpt Square fill Bsquare } bind def
/S14 { BL [] 0 setdash 2 copy exch vpt sub exch vpt sub vpt2 vpt Rec fill
       2 copy exch vpt sub exch vpt Square fill Bsquare } bind def
/S15 { BL [] 0 setdash 2 copy Bsquare fill Bsquare } bind def
/D0 { gsave translate 45 rotate 0 0 S0 stroke grestore } bind def
/D1 { gsave translate 45 rotate 0 0 S1 stroke grestore } bind def
/D2 { gsave translate 45 rotate 0 0 S2 stroke grestore } bind def
/D3 { gsave translate 45 rotate 0 0 S3 stroke grestore } bind def
/D4 { gsave translate 45 rotate 0 0 S4 stroke grestore } bind def
/D5 { gsave translate 45 rotate 0 0 S5 stroke grestore } bind def
/D6 { gsave translate 45 rotate 0 0 S6 stroke grestore } bind def
/D7 { gsave translate 45 rotate 0 0 S7 stroke grestore } bind def
/D8 { gsave translate 45 rotate 0 0 S8 stroke grestore } bind def
/D9 { gsave translate 45 rotate 0 0 S9 stroke grestore } bind def
/D10 { gsave translate 45 rotate 0 0 S10 stroke grestore } bind def
/D11 { gsave translate 45 rotate 0 0 S11 stroke grestore } bind def
/D12 { gsave translate 45 rotate 0 0 S12 stroke grestore } bind def
/D13 { gsave translate 45 rotate 0 0 S13 stroke grestore } bind def
/D14 { gsave translate 45 rotate 0 0 S14 stroke grestore } bind def
/D15 { gsave translate 45 rotate 0 0 S15 stroke grestore } bind def
/DiaE { stroke [] 0 setdash vpt add M
  hpt neg vpt neg V hpt vpt neg V
  hpt vpt V hpt neg vpt V closepath stroke } def
/BoxE { stroke [] 0 setdash exch hpt sub exch vpt add M
  0 vpt2 neg V hpt2 0 V 0 vpt2 V
  hpt2 neg 0 V closepath stroke } def
/TriUE { stroke [] 0 setdash vpt 1.12 mul add M
  hpt neg vpt -1.62 mul V
  hpt 2 mul 0 V
  hpt neg vpt 1.62 mul V closepath stroke } def
/TriDE { stroke [] 0 setdash vpt 1.12 mul sub M
  hpt neg vpt 1.62 mul V
  hpt 2 mul 0 V
  hpt neg vpt -1.62 mul V closepath stroke } def
/PentE { stroke [] 0 setdash gsave
  translate 0 hpt M 4 {72 rotate 0 hpt L} repeat
  closepath stroke grestore } def
/CircE { stroke [] 0 setdash 
  hpt 0 360 arc stroke } def
/Opaque { gsave closepath 1 setgray fill grestore 0 setgray closepath } def
/DiaW { stroke [] 0 setdash vpt add M
  hpt neg vpt neg V hpt vpt neg V
  hpt vpt V hpt neg vpt V Opaque stroke } def
/BoxW { stroke [] 0 setdash exch hpt sub exch vpt add M
  0 vpt2 neg V hpt2 0 V 0 vpt2 V
  hpt2 neg 0 V Opaque stroke } def
/TriUW { stroke [] 0 setdash vpt 1.12 mul add M
  hpt neg vpt -1.62 mul V
  hpt 2 mul 0 V
  hpt neg vpt 1.62 mul V Opaque stroke } def
/TriDW { stroke [] 0 setdash vpt 1.12 mul sub M
  hpt neg vpt 1.62 mul V
  hpt 2 mul 0 V
  hpt neg vpt -1.62 mul V Opaque stroke } def
/PentW { stroke [] 0 setdash gsave
  translate 0 hpt M 4 {72 rotate 0 hpt L} repeat
  Opaque stroke grestore } def
/CircW { stroke [] 0 setdash 
  hpt 0 360 arc Opaque stroke } def
/BoxFill { gsave Rec 1 setgray fill grestore } def
end
}
\begin{picture}(3600,2160)(0,0)
\special{"
gnudict begin
gsave
0 0 translate
0.100 0.100 scale
0 setgray
newpath
1.000 UL
LTb
350 200 M
63 0 V
3087 0 R
-63 0 V
350 510 M
63 0 V
3087 0 R
-63 0 V
350 820 M
63 0 V
3087 0 R
-63 0 V
350 1130 M
63 0 V
3087 0 R
-63 0 V
350 1440 M
63 0 V
3087 0 R
-63 0 V
350 1750 M
63 0 V
3087 0 R
-63 0 V
350 2060 M
63 0 V
3087 0 R
-63 0 V
350 200 M
0 63 V
0 1797 R
0 -63 V
666 200 M
0 63 V
0 1797 R
0 -63 V
851 200 M
0 63 V
0 1797 R
0 -63 V
982 200 M
0 63 V
0 1797 R
0 -63 V
1084 200 M
0 63 V
0 1797 R
0 -63 V
1167 200 M
0 63 V
0 1797 R
0 -63 V
1237 200 M
0 63 V
0 1797 R
0 -63 V
1298 200 M
0 63 V
0 1797 R
0 -63 V
1352 200 M
0 63 V
0 1797 R
0 -63 V
1400 200 M
0 63 V
0 1797 R
0 -63 V
1716 200 M
0 63 V
0 1797 R
0 -63 V
1901 200 M
0 63 V
0 1797 R
0 -63 V
2032 200 M
0 63 V
0 1797 R
0 -63 V
2134 200 M
0 63 V
0 1797 R
0 -63 V
2217 200 M
0 63 V
0 1797 R
0 -63 V
2287 200 M
0 63 V
0 1797 R
0 -63 V
2348 200 M
0 63 V
0 1797 R
0 -63 V
2402 200 M
0 63 V
0 1797 R
0 -63 V
2450 200 M
0 63 V
0 1797 R
0 -63 V
2766 200 M
0 63 V
0 1797 R
0 -63 V
2951 200 M
0 63 V
0 1797 R
0 -63 V
3082 200 M
0 63 V
0 1797 R
0 -63 V
3184 200 M
0 63 V
0 1797 R
0 -63 V
3267 200 M
0 63 V
0 1797 R
0 -63 V
3337 200 M
0 63 V
0 1797 R
0 -63 V
3398 200 M
0 63 V
0 1797 R
0 -63 V
3452 200 M
0 63 V
0 1797 R
0 -63 V
3500 200 M
0 63 V
0 1797 R
0 -63 V
1.000 UL
LTb
350 200 M
3150 0 V
0 1860 V
-3150 0 V
350 200 L
1.000 UL
LT0
3137 1947 M
263 0 V
350 1515 M
13 0 V
21 0 V
21 0 V
20 0 V
21 0 V
20 0 V
21 0 V
21 0 V
20 0 V
21 0 V
20 1 V
21 0 V
21 0 V
20 0 V
21 0 V
20 0 V
21 0 V
21 0 V
20 0 V
21 0 V
20 0 V
21 0 V
20 0 V
21 0 V
21 0 V
20 0 V
21 0 V
20 0 V
21 0 V
21 0 V
20 0 V
21 0 V
20 0 V
21 0 V
21 0 V
20 0 V
21 0 V
20 0 V
21 0 V
21 -1 V
20 0 V
21 0 V
20 0 V
21 -1 V
21 0 V
20 -1 V
21 -1 V
20 -1 V
21 -2 V
21 -1 V
20 -3 V
21 -2 V
20 -3 V
21 -4 V
21 -4 V
20 -6 V
21 -6 V
20 -8 V
21 -9 V
20 -10 V
21 -12 V
21 -13 V
20 -16 V
21 -17 V
20 -19 V
21 -20 V
21 -22 V
20 -23 V
21 -24 V
20 -24 V
21 -24 V
21 -24 V
20 -23 V
21 -23 V
20 -21 V
21 -21 V
21 -20 V
20 -19 V
21 -18 V
20 -19 V
21 -17 V
21 -18 V
20 -17 V
21 -17 V
20 -16 V
21 -16 V
21 -16 V
20 -16 V
21 -16 V
20 -16 V
21 -15 V
20 -15 V
21 -14 V
21 -14 V
20 -14 V
21 -14 V
20 -13 V
21 -13 V
21 -12 V
20 -11 V
21 -10 V
20 -10 V
21 -9 V
21 -9 V
20 -7 V
21 -7 V
20 -6 V
21 -6 V
21 -4 V
20 -4 V
21 -3 V
20 -2 V
21 -1 V
21 -2 V
20 0 V
21 0 V
20 0 V
21 2 V
21 1 V
20 3 V
21 0 V
20 3 V
21 2 V
21 2 V
20 3 V
21 1 V
20 2 V
21 1 V
20 0 V
21 1 V
21 1 V
20 -1 V
21 -1 V
20 -1 V
21 -1 V
21 -2 V
20 -2 V
21 -2 V
20 -3 V
21 -2 V
21 -4 V
20 -2 V
21 -6 V
20 1 V
21 -5 V
21 0 V
20 -8 V
21 -2 V
20 -9 V
21 5 V
21 -10 V
20 6 V
21 -5 V
6 0 V
1.000 UL
LT1
3137 1847 M
263 0 V
350 1452 M
13 0 V
21 0 V
21 0 V
20 0 V
21 0 V
20 0 V
21 0 V
21 0 V
20 1 V
21 0 V
20 0 V
21 0 V
21 0 V
20 0 V
21 0 V
20 0 V
21 1 V
21 0 V
20 0 V
21 0 V
20 0 V
21 1 V
20 0 V
21 0 V
21 1 V
20 0 V
21 0 V
20 1 V
21 0 V
21 1 V
20 0 V
21 1 V
20 1 V
21 0 V
21 1 V
20 1 V
21 1 V
20 1 V
21 1 V
21 1 V
20 1 V
21 1 V
20 1 V
21 2 V
21 1 V
20 1 V
21 2 V
20 1 V
21 2 V
21 1 V
20 1 V
21 1 V
20 1 V
21 0 V
21 0 V
20 -1 V
21 -1 V
20 -3 V
21 -3 V
20 -5 V
21 -6 V
21 -7 V
20 -9 V
21 -10 V
20 -12 V
21 -12 V
21 -14 V
20 -14 V
21 -14 V
20 -14 V
21 -13 V
21 -12 V
20 -12 V
21 -10 V
20 -9 V
21 -9 V
21 -8 V
20 -8 V
21 -8 V
20 -8 V
21 -7 V
21 -8 V
20 -6 V
21 -6 V
20 -6 V
21 -6 V
21 -6 V
20 -7 V
21 -5 V
20 -5 V
21 -5 V
20 -4 V
21 -4 V
21 -4 V
20 -3 V
21 -2 V
20 -2 V
21 -1 V
21 -1 V
20 0 V
21 0 V
20 1 V
21 0 V
21 1 V
20 2 V
21 1 V
20 1 V
21 1 V
21 1 V
20 1 V
21 1 V
20 1 V
21 1 V
21 1 V
20 1 V
21 2 V
20 2 V
21 2 V
21 2 V
20 1 V
21 2 V
20 1 V
21 1 V
21 0 V
20 0 V
21 0 V
20 0 V
21 0 V
20 0 V
21 1 V
21 0 V
20 1 V
21 0 V
20 0 V
21 -2 V
21 0 V
20 0 V
21 0 V
20 0 V
21 0 V
21 0 V
20 0 V
21 0 V
20 0 V
21 0 V
21 0 V
20 0 V
21 0 V
20 -1 V
21 1 V
21 0 V
20 0 V
21 0 V
6 -1 V
1.000 UL
LT2
3137 1747 M
263 0 V
350 1384 M
13 0 V
21 0 V
21 0 V
20 0 V
21 0 V
20 0 V
21 0 V
21 0 V
20 1 V
21 0 V
20 0 V
21 0 V
21 0 V
20 0 V
21 0 V
20 0 V
21 0 V
21 0 V
20 1 V
21 0 V
20 0 V
21 0 V
20 0 V
21 1 V
21 0 V
20 0 V
21 1 V
20 0 V
21 0 V
21 1 V
20 0 V
21 1 V
20 0 V
21 1 V
21 1 V
20 0 V
21 1 V
20 1 V
21 1 V
21 1 V
20 1 V
21 1 V
20 1 V
21 2 V
21 1 V
20 2 V
21 1 V
20 2 V
21 2 V
21 2 V
20 2 V
21 2 V
20 2 V
21 2 V
21 2 V
20 2 V
21 2 V
20 2 V
21 2 V
20 1 V
21 0 V
21 1 V
20 -1 V
21 -2 V
20 -2 V
21 -3 V
21 -5 V
20 -5 V
21 -7 V
20 -7 V
21 -9 V
21 -8 V
20 -9 V
21 -10 V
20 -9 V
21 -9 V
21 -8 V
20 -9 V
21 -9 V
20 -9 V
21 -9 V
21 -9 V
20 -9 V
21 -10 V
20 -9 V
21 -9 V
21 -10 V
20 -9 V
21 -10 V
20 -10 V
21 -10 V
20 -10 V
21 -10 V
21 -10 V
20 -10 V
21 -11 V
20 -10 V
21 -10 V
21 -10 V
20 -10 V
21 -10 V
20 -9 V
21 -10 V
21 -9 V
20 -9 V
21 -9 V
20 -9 V
21 -8 V
21 -9 V
20 -8 V
21 -8 V
20 -8 V
21 -8 V
21 -7 V
20 -8 V
21 -7 V
20 -8 V
21 -7 V
21 -7 V
20 -7 V
21 -7 V
20 -7 V
21 -6 V
21 -7 V
20 -6 V
21 -7 V
20 -6 V
21 -6 V
20 -6 V
21 -6 V
21 -5 V
20 -6 V
21 -5 V
20 -6 V
21 -5 V
21 -5 V
20 -5 V
21 -5 V
20 -5 V
21 -5 V
21 -4 V
20 -5 V
21 -5 V
20 -4 V
21 -4 V
21 -4 V
20 -5 V
21 -4 V
20 -3 V
21 -4 V
21 -4 V
20 -4 V
21 -3 V
6 -2 V
1.000 UL
LT3
350 1440 M
32 0 V
32 0 V
31 0 V
32 0 V
32 0 V
32 0 V
32 0 V
32 0 V
31 0 V
32 0 V
32 0 V
32 0 V
32 0 V
31 0 V
32 0 V
32 0 V
32 0 V
32 0 V
32 0 V
31 0 V
32 0 V
32 0 V
32 0 V
32 0 V
31 0 V
32 0 V
32 0 V
32 0 V
32 0 V
32 0 V
31 0 V
32 0 V
32 0 V
32 0 V
32 0 V
31 0 V
32 0 V
32 0 V
32 0 V
32 0 V
32 0 V
31 0 V
32 0 V
32 0 V
32 0 V
32 0 V
31 0 V
32 0 V
32 0 V
32 0 V
32 0 V
32 0 V
31 0 V
32 0 V
32 0 V
32 0 V
32 0 V
31 0 V
32 0 V
32 0 V
32 0 V
32 0 V
32 0 V
31 0 V
32 0 V
32 0 V
32 0 V
32 0 V
31 0 V
32 0 V
32 0 V
32 0 V
32 0 V
32 0 V
31 0 V
32 0 V
32 0 V
32 0 V
32 0 V
31 0 V
32 0 V
32 0 V
32 0 V
32 0 V
32 0 V
31 0 V
32 0 V
32 0 V
32 0 V
32 0 V
31 0 V
32 0 V
32 0 V
32 0 V
32 0 V
32 0 V
31 0 V
32 0 V
32 0 V
stroke
grestore
end
showpage
}
\put(3087,1747){\makebox(0,0)[r]{${\cal R}_{F_2}$}}
\put(3087,1847){\makebox(0,0)[r]{${\cal R}_v$}}
\put(3087,1947){\makebox(0,0)[r]{${\cal R}_g$}}
\put(666,1750){\makebox(0,0)[l]{{\Large $^{40}$Ca}}}
\put(0,1130){%
\special{ps: gsave currentpoint currentpoint translate
270 rotate neg exch neg exch translate}%
\makebox(0,0)[b]{\shortstack{${\cal R}$}}%
\special{ps: currentpoint grestore moveto}%
}
\put(300,2060){\makebox(0,0)[r]{1.2}}
\put(300,1750){\makebox(0,0)[r]{1.1}}
\put(300,1440){\makebox(0,0)[r]{1}}
\put(300,1130){\makebox(0,0)[r]{0.9}}
\put(300,820){\makebox(0,0)[r]{0.8}}
\put(300,510){\makebox(0,0)[r]{0.7}}
\put(300,200){\makebox(0,0)[r]{0.6}}
\end{picture}